# Comparison among Klein-Gordon equation, Dirac equation and relativistic stationary Schrödinger equation

Guang-jiong Ni
*Department of Physics, Fudan University,*
*Shanghai, 200433, P. R. China*
*E-mail: Gjni@fudan.ac.cn*

Weimin Zhou and Jun Yan
*Department of Physics, New York University, 4 Washington Place,*
*New York, NY, 10003, USA*
*E-mail :wz214@is6.nyu.edu and jy272@scires.nyu.edu*

**Abstract**

A particle is always not pure. It always contains hiding antiparticle ingredient which is the essence of special relativity. Accordingly, the Klein-Gordon (KG) equation and Dirac equation are restudied and compared with the Relativistic Stationary Schrödinger Equation (RSSE). When an electron is bound in a Hydrogenlike atom with pointlike nucleus having charge number $Z$, the critical value of $Z$, $Z_c$, equals to 137 in Dirac equation whereas $Z_c = \sqrt{\frac{M}{\mu}}(137)$ in RSSE with $M$ and $\mu$ being the total mass of atom and the reduced mass of the electron.

## 1 Introduction

The Einstein mass-energy relation $E = mc^2$ reveals the simple proportionality between energy $E$ and mass $m$ of any matter with $c$ being the speed of light. For a free particle moving with velocity $v$, the mass $m$ is related to its rest mass $m_0$ as $m = \frac{m_0}{\sqrt{1-(\frac{v}{c})^2}}$, which approaches infinity when $v$ approaches to $c$. On the other hand, when an electron is bound in the Coulomb field of point nucleus with charge $Ze$ ($e > 0$) to form a hydrogenlike atom, according to the prediction of Dirac equation (with infinite nucleus mass $m_N \to \infty$), the electron energy will decrease. For $1S$ state it will approach to zero when $Z$ approaches 137.

The aim of this paper is to show that all the above variation in particle mass can be ascribed to the variation of relative ratio of hiding antimatter to matter in a particle. That is the ratio $R$ of hiding positron ingredient to electron ingredient in a electron, which determines the mass of electron in free motion case and the total energy $E$ of Hydrogenlike atom in binding case. While there is a upper bound for the velocity of electron $v_{\max} = c$, ($m_{\max} = \infty$), there is also a lower bound for the total energy $E$ in the later case $E_{\min} = 0$ at $Z_c = \sqrt{\frac{M}{\mu}}(\frac{1}{\alpha}) = \sqrt{\frac{M}{\mu}}(137)$ with $M$ and $\mu$ being the total rest mass ($M = m_e + m_N$) and reduced mass $\mu = \frac{m_e m_N}{M}$. Both the upper and lower bounds are fixed by the condition $R = 1$.

The organization of this paper is as follows. In Section II and III the Klein-Gordon (KG) equation and Dirac equation are discussed respectively to show the same role of symmetry of particle-antiparticle played in the relativistic quantum mechanics but with different outcome. In particular, the critical value $Z_c$ for KG equation or Dirac equation is $\frac{1}{2\alpha} = \frac{1}{2}(137)$ or $\frac{1}{\alpha} = 137$ respectively. Being an improvement of one-body equation, the two-body relativistic stationary Schrödinger equation (RSSE) is discussed further in section III to derive a much larger value $Z_c = \sqrt{\frac{M}{\mu}}(137)$. The experimental implication of this improvement and the relevant problems are discussed at the final section V.

## 2 Klein-Gordon equation

For simplicity, we begin from a particle with mass $m_0$ but without spin. It is described in nonrelativistic quantum mechanics by the Schrödinger equation. Then its kinetic energy reads $\frac{1}{2}m_0 v^2$ with velocity $v$ being unlimited and $m_0$ unchanged. When it carries a charge $(-e)$ and is bound in the Coulomb field with potential energy

$$V(r) = -\frac{Ze^2}{4\pi\varepsilon_0 r} \tag{1}$$

The binding energy $B$ is well known as ($\alpha = \frac{e^2}{4\pi\varepsilon_0\hbar c} \simeq \frac{1}{137}$ with $\hbar$ being the Plank constant)

$$B = \frac{Z^2\alpha^2}{2n^2} m_0 c^2 \qquad (n = 1, 2, \cdots) \tag{2}$$

In other words, the mass of electron $m = m_0 - \frac{B}{c^2}$ would decrease without lower bound if the charge number of nucleus $Z$ is sufficiently large.

However, the situation becomes quite different in the theory of special relativity. Consider a meson $\pi^-$ bound in a point nucleus (with infinite mass $m_N \to \infty$). Its wave function $\psi(\vec{x}, t)$ satisfies the Klein-Gordon (K-G) equation. [1,2]

$$(i\hbar\frac{\partial}{\partial t} - V(r))^2\psi = m_0^2 c^4 \psi - c^2\hbar^2\nabla^2\psi \tag{3}$$

According to the pioneer work of Feshbach and Villars [3], the main point of view in this paper is as follows. We should look at $\psi$ being composed of two kinds of fields

$$\begin{cases} \varphi &= \frac{1}{2}[(1 - \frac{V}{m_0c^2})\psi + i\frac{\hbar}{m_0c^2}\dot{\psi}] \\ \chi &= \frac{1}{2}[(1 + \frac{V}{m_0c^2})\psi - i\frac{\hbar}{m_0c^2}\dot{\psi}] \end{cases} \tag{4}$$

Then Eq.(3) can be recast into the form of coupling Schrödinger equations:

$$\begin{cases} (i\hbar\frac{\partial}{\partial t} - V)\varphi & = m_0 c^2 \varphi - \frac{\hbar^2}{2m_0}\nabla^2(\varphi + \chi) \\ (i\hbar\frac{\partial}{\partial t} - V)\chi & = -m_0 c^2 \chi + \frac{\hbar^2}{2m_0}\nabla^2(\chi + \varphi) \end{cases} \tag{5}$$

Eq.(5) is invariant under the transformation $(\vec{x} \to -\vec{x}, t \to -t)$ and

$$\varphi(-\vec{x}, -t) \quad \to \quad \chi(\vec{x}, t) \tag{6}$$

$$V(-\vec{x}, -t) \quad \to \quad -V(\vec{x}, t) \tag{7}$$

The meaning of $\varphi$ and $\chi$ can be seen from the continuity equation:

$$\frac{\partial \rho}{\partial t} + \nabla \cdot \vec{j} = 0 \tag{8}$$

with the "probability density"

$$\rho = |\varphi|^2 - |\chi|^2 = \varphi^* \varphi - \chi^* \chi \tag{9}$$

and the "current density"

$$\begin{aligned} \vec{j} = \tfrac{i\hbar}{2m_0}[&(\varphi\nabla\varphi^* - \varphi^*\nabla\varphi) + (\chi\nabla\chi^* - \chi^*\nabla\chi) \\ &+ (\varphi\nabla\chi^* - \chi^*\nabla\varphi) + (\chi\nabla\varphi^* - \varphi^*\nabla\chi)] \end{aligned} \tag{10}$$

We explain the field $\varphi$ being the "particle (matter) ingredient" of a particle, whereas $\chi$ being the hiding "antiparticle (antimatter) ingredient" inside a particle.

See first the free motion case $V = 0$. The particle is described by a plane wave function along $z$ axis:

$$\psi \sim \exp\left\{\frac{i}{\hbar}(pz - Et)\right\} \tag{11}$$

Beginning from $E = m_0 c^2$, $|\chi|$ increases from zero until the limit of momentum $p \to \infty$, i.e., $E \to \infty$, or

$$\lim_{v \to c} |\chi| \to |\varphi| \tag{12}$$

Let us discuss the wave packet:

$$\psi(z,t) = \int_{-\infty}^{\infty} (\frac{\sigma}{\pi})^{\frac{1}{4}} e^{-\frac{k^2}{2\sigma}} e^{i(kz - \omega t)} dk \tag{13}$$

with $\hbar\omega = \sqrt{\hbar^2 k^2 c^2 + m_0^2 c^4} \simeq m_0 c^2 + \frac{\hbar^2 k^2}{2m_0} + \cdots$

Assume $\sqrt{\sigma} \ll \frac{m_0 c}{\hbar}$, then

$$\psi(z,t) \simeq \frac{(\frac{\sigma}{\pi})^{\frac{1}{4}}}{(1+\frac{i\sigma\hbar t}{m_0})^{\frac{1}{2}}} \exp\left\{-\frac{\sigma z^2}{2(1+\frac{i\sigma\hbar t}{m_0})} - \frac{i m_0 c^2 t}{\hbar}\right\} \tag{14}$$

If consider $\frac{\sigma\hbar t}{m_0} \ll 1$ to ignore the spreading of wave packet in low speed case ($v \ll c$). Then we perform a "boost" transformation, i.e., to push the wave packet to high speed ($v \to c$) case. Thus we see in the figure 1 that:

(i) The width of packet shrinks — Lorentz contraction.

(ii) The amplitude of $\rho$ increases — "boost" effect.

(iii) The new observation is that both $|\varphi|^2$ and $|\chi|^2$ in $\rho$ increase even more sharply while keeping $|\varphi| > |\chi|$ to preserve $|\varphi + \chi| \sim |\psi|$ invariant.

The ratio of hiding $|\chi|^2$ to $|\varphi|^2$ reads:

$$R^{KG}_{free} = \frac{\int_{-\infty}^{\infty} |\chi|^2 dz}{\int_{-\infty}^{\infty} |\varphi|^2 dz} = \left[\frac{1-\sqrt{1-(\frac{v}{c})^2}}{1+\sqrt{1-(\frac{v}{c})^2}}\right]^2 \tag{15}$$

It is interesting to see the stationary $1S$ state (zero angular momentum state with principal quantum number $n = 1$) in field $V(r)$ shown in Eq.(1).

Now the energy level is quantized to be

$$E^{KG}_{1S} = m_0 c^2 \sqrt{\frac{1}{2} + \sqrt{\frac{1}{4} - Z^2\alpha^2}} \tag{16}$$

which is a function of Z. When $Z \to \frac{1}{2\alpha} \simeq \frac{137}{2}$, the energy $E^{KG}_{1S}$ decreases to a lowest limit $\frac{m_0 c^2}{\sqrt{2}}$. Meanwhile, the ratio

$$R^{KG}_{1S} = \frac{\int |\chi|^2 d\vec{x}}{\int |\varphi|^2 d\vec{x}} = 1 - 4\left[2 + \left(y + \frac{1}{2}\right)^{\frac{1}{2}} + \frac{\left(y+\frac{1}{2}\right)^{\frac{3}{2}}}{2y}\right]^{-1} \tag{17}$$

$y = \sqrt{\frac{1}{4} - Z^2\alpha^2}$, increases from zero to the upper limit 1, as shown in the figure 2.

## 3 Dirac equation

Next turn to the electron case. Being a particle with spin $\frac{1}{2}$, it is described by a Dirac spinor wave function

$$\Psi = \begin{pmatrix} \varphi \\ \chi \end{pmatrix} \tag{18}$$

with four components. Here $\varphi$ and $\chi$, (each with two components) usually called as the "positive" and "negative" energy components in the literature [1], [2] are just the counterpart of $\varphi$ and $\chi$ for particle without spin.

However, in this case, instead of (9), we have

$$\rho_{Dirac} = \Psi^\dagger \Psi = \varphi^\dagger \varphi + \chi^\dagger \chi \tag{19}$$

Hence for a freely moving electron wave packet, instead of figure 1, we have figure 3. One sees that both $\varphi^\dagger\varphi$ and $\chi^\dagger\chi$ are increasing with the velocity $v$.

But they are constrained within the boosting $\rho$ and the invariant quantity during the boosting process is

$$\overline{\Psi}\Psi = \varphi^\dagger \varphi - \chi^\dagger \chi > 0 \tag{20}$$

where the inequality ensures that the electron is always an electron though the hiding "antielectron (positron)" ingredient $|\chi|$ is already approaching $|\varphi|$ when $v \to c$. The ratio reads

$$R_{free}^{Dirac} = \frac{1 - \sqrt{1 - \left(\frac{v}{c}\right)^2}}{1 + \sqrt{1 - \left(\frac{v}{c}\right)^2}} \tag{21}$$

On the other hand, when the electron is bound inside a hydrogenlike atom, the energy level of $1S$ state is

$$E_{1S}^{Dirac} = m_0 c^2 \sqrt{1 - Z^2 \alpha^2} \tag{22}$$

which decreases to zero when $Z \to \frac{1}{\alpha} \simeq 137$ as shown in figure 4. Meanwhile the ratio

$$R_{1S}^{Dirac} = \frac{1 - \sqrt{1 - Z^2\alpha^2}}{1 + \sqrt{1 - Z^2\alpha^2}} \tag{23}$$

increases from zero to 1 , similar to Eq.(17) and the curve in figure 2.

## 4 Relativistic Stationary Schrödinger equation (RSSE)

As is well known, the one-body KG equation and Dirac equation are derived in covariant formalism, i.e., by the combination of the principle of special relativity (SR) with that of quantum mechanics. However, it is not easy to generalize them into two-body or many-body case. Now the analysis in above two sections show that the principle of SR is equivalent to the coexistence of $\varphi$ and $\chi$ states in a particle with the symmetry of transmutation between them under the space-time inversion ($\vec{x} \to -\vec{x}, t \to -t$) as shown at the Eq.(6).[4,5] So we manage to establish an equation for two-particle system based on this symmetry. Denote the coordinates and masses of two particles by $\vec{r_1}$, $\vec{r_2}$ and $m_1$, $m_2$. The particle and antiparticle ingredients are described by wave functions $\varphi(\vec{r_1}, \vec{r_2}, t)$ and $\chi(\vec{r_1}, \vec{r_2}, t)$. We propose the coupling equation as follows ($\hbar = c = 1$):

$$\left\{ \begin{array}{lcll} i\frac{\partial\varphi}{\partial t} & = & (m_1+m_2)\varphi & +V(|\vec{r_1}-\vec{r_2}|)(\varphi+\chi) \\ & & -\frac{1}{2m_1}\nabla_1^2(\varphi+\chi) & -\frac{1}{2m_2}\nabla_2^2(\varphi+\chi) \\ i\frac{\partial\chi}{\partial t} & = & -(m_1+m_2)\chi & -V(|\vec{r_1}-\vec{r_2}|)(\varphi+\chi) \\ & & +\frac{1}{2m_1}\nabla_1^2(\varphi+\chi) & +\frac{1}{2m_2}\nabla_2^2(\varphi+\chi) \end{array} \right. \tag{24}$$

Note that, however, here $V(-\vec{r_1}, -\vec{r_2}, -t) = V(\vec{r_1}, \vec{r_2}, t)$ in contrast with the Eq.(7). The reason lies on the dynamical nature of nucleus (particle 1) which would transform into antinucleus under space-time inversion whereas the $V$ in Eq.(7) is treated as an external potential imposed by the inert core without change in space-time inversion. Furthermore, here both $\varphi$ and $\chi$ is involved in the term containing $V$ in contrast with Eq.(5). Of course, these two points can only be verified by the equation derived and eventually by experiment.[6]

Introducing the coordinate of center of mass $\vec{R}_c = \frac{1}{M}(m_1\vec{r_1} + m_2\vec{r_2})$ and the relative coordinate $\vec{r} = \vec{r_2} - \vec{r_1}$, we derive from Eq.(24):

$$\left\{ \begin{array}{l} -\frac{1}{M}\frac{\partial^2(\varphi+\chi)}{\partial t^2} = M(\varphi+\chi) + 2V(\varphi+\chi) - \frac{1}{\mu}\nabla_r^2(\varphi+\chi) - \frac{1}{M}\nabla_R^2(\varphi+\chi) \\ (\varphi-\chi) = \frac{i}{M}\frac{\partial(\varphi+\chi)}{\partial t} \end{array} \right. \tag{25}$$

Considering a stationary state with momentum $\vec{P}$ in the system of center of mass, we set

$$(\varphi+\chi) = \psi(\vec{r}) \exp\left\{ \frac{i}{\hbar}(\vec{P}\cdot\vec{R} - Et) \right\} \tag{26}$$

where $E$ is the total energy of system, yielding

$$\left\{ \begin{array}{l} -\frac{1}{2\mu}\nabla_r^2\psi(\vec{r}) + V(\vec{r})\psi(\vec{r}) = \varepsilon\psi(\vec{r}) \\ \varepsilon = \frac{1}{2M}(E^2 - M^2 - P^2) \end{array} \right. \tag{27}$$

For a binding state, the binding energy $B$ of system is defined as ($\vec{P} = 0$) $B = M - E$ so

$$B = M\left[1 - (1 + \frac{2\varepsilon}{M})^{1/2}\right] \tag{28}$$

when $\varepsilon/M \ll 1$, $B \simeq -\varepsilon$ as expected.

Eq.(27) shows that the stationary Schrödinger equation is essentially relativistic as long as its eigenvalue $\varepsilon$ being not directly $(-B)$ but indirectly related to $B$ via Eq.(28).

This RSSE has an important feature that its eigenvalue $\varepsilon$ has a lower bound $\varepsilon_{\min} = -M/2$ as shown in Eq.(28), corresponding to $B_{\max} = M$ or $E_{\min} = 0$. Just like in the case of KG or Dirac equation, this lower bound is also characterized by the condition. $\lim_{E\to 0}\chi = \varphi$ as shown by Eq.(25). But this time it leads to the critical value of $Z$ for a Hydrogenlike atom being

$$Z_c = \sqrt{\frac{M}{\mu}}\frac{1}{\alpha} = \sqrt{\frac{M}{\mu}}(137) \tag{29}$$

which is much larger than that in KG or Dirac equation since $\frac{M}{\mu} \gg 1$. This is no surprise because of the different meaning of $V$ discussed above. Actually the concept of binding energy $B$ is related to the whole system composed of two particles, so in Eq.(29) two mass parameters $M$ and $\mu$ are involved. To treat the nucleus as an inert core is not so reasonable. We would like to stress that in RSSE case there is no any singularity at $B = m_2 = m_e$ (when $Z_c' \simeq \sqrt{2}(137)$) or $B = 2m_e$ (when $Z_c'' \simeq 2(137)$) where $|\chi| \ll |\varphi|$ still. No considerable instability of atom could be expected at $Z_c'$ or $Z_c''$. Being a onebody equation, the Dirac equation, in our point of view, had overestimated the antiparticle ingredient of an electron in the hydrogenlike atom. On the other hand, the RSSE may be not applicable to the case of real hydrogenlike atom because of $m_N \gg m_e$ $(m \gg \mu)$ condition. However, it works well for the heavy quarkonium model case $(m_q = m_{\bar{q}})$, see Ref. [6].

## 5 Summary and discussion

(a) In our point of view, there is no any "negative energy electron sea" filling the continuum states ranging from $(-m_ec^2)$ to $(-\infty)$. If defining unrigorously the electron energy $E_e = E - B - m_Nc^2$ in the Hydrogenlike atom (with extended nucleus having $Z \approx 172$), when the energy of 1$S$ state decreases until $E_e < -m_ec^2$, it was often said that the electron is diving into the "sea" and may trigger the emission of positron and the formation of "charged vacuum". To our knowledge, this kind of concept provided a tentative explanation to the so-called GSI "$e^+e^-$ puzzle" in the experiment of heavy ion collision. However, there is a strong doubt about the experimental accuracy.[7] We are far from the experts in this field. But theoretically, we tend to believe that no considerable instability of nuclei could be expected at the present experiments because of our high estimation of critical value of Eq.(29).

(b) In nonrelativistic quantum mechanics only the particle (e.g., the electron) is considered. the velocity of particle can enhance without a upper limit. On the other hand, the energy of a binding particle can decrease without a lower limit either. Its mass $m_0$ remains unchanged in any case.

(c) In relativistic quantum mechanics, a particle is always not pure. It is accompanied by its hiding antiparticle ingredient essentially. If a free rest particle with mass $m_0$ described by $\varphi(\vec{x},t)$, the accompanying $\chi(\vec{x},t)$ will be excited coherently once the particle is set into motion or bound in a system. Then its velocity $v$ is bound from above by a limiting speed $c$ ($E_{\max}=\infty$) while its energy $E$ of bound system is bound from below: $E_{\min}=0$. At both sides ($E=\infty$ or 0), the ratio of ingredient of antiparticle to that of particle: $R=\int|\chi|^2 d\vec{x}/\int|\varphi|^2 d\vec{x}\to 1$.

(d) The common essence of any matter is the basic symmetry Eq. (6). It could be stated as a postulate that "the space-time inversion ($\vec{x}\to-\vec{x}$, $t\to-t$) is equivalent to the transformation between particle and antiparticle".[4,5]

(e) However, inside a particle, $\varphi$ always dominates $\chi$, i.e., $|\varphi|>|\chi|$. So they do not exhibit the symmetry Eq.(6) explicitly. Being the "slave" in the particle, $\chi$ has to obey the "master" $\varphi$. In particular, the wave function for an electron in freely motion reads always as

$$\Psi_{e^-}\sim\varphi\sim\chi\sim\exp\left\{\frac{i}{\hbar}(\vec{p}\cdot\vec{x}-E\cdot t)\right\},\qquad(|\varphi|>|\chi|)\tag{30}$$

On the other hand, if we perform a space-time inversion, $\varphi(\vec{x},t)\to\varphi(-\vec{x},-t)=\chi_c(\vec{x},t)$ becomes the "master", whereas $\chi(\vec{x},t)\to\chi(-\vec{x},-t)=\varphi_c(\vec{x},t)$ reduces into the slave. Then Eq.(30) turns into the wave function for a positron:

$$\Psi_{e^+}\sim\chi_c\sim\varphi_c\sim\exp\left\{-\frac{i}{\hbar}(\vec{p}\cdot\vec{x}-E\cdot t)\right\},\qquad(|\chi_c|>|\varphi_c|)\tag{31}$$

with the same momentum $\vec{p}$ and energy $E(>0)$. The observation of Eqs. (30) and (31) was put forward quite early by Schwinger [8], Konopinski and Mahmand [9] and even essentially by Stüeckelberg [10] and Feynman [11]. See also Refs. [12] and [13].

(f) The ratio $R<1$ could be viewed as an order parameter characterizing the status of a "particle". Formally, if we always define $R=\int|\chi|^2 d\vec{x}/\int|\varphi|^2 d\vec{x}$, then $R>1$ will characterize the status of an "antiparticle". In other words, we look at the "negative energy" state of a particle directly as the "positive energy" state of its antiparticle, either for KG particle or for Dirac particle. It seems to us that the historical mission of the concept of hole theory for electron is coming to an end.

(g) Actually, all the strange effects (including the Lorentz transformation) in special relativity can be derived by the symmetry Eq.(6) in combination with the principles of quantum mechanics.[4,5,6]

(h) For further discussion on the concept in contemporary physics, see Ref. [14].

**Acknowledgments**

We thank Profs. R. Brandt, J. J. Griffin, and D. Zwanziger for discussions. We also thank Mr. Sangtian Liu in NYU for giving us a lot of help in LaTeXand figures of this paper. This work was supported in part by the NSF in China.

**References**

1. J. D. Bjorken and S. D. Drell, *Relativistic Quantum Mechanics*, McGraw-Hill Book Company, 1964.
2. J. J. Sakurai, *Advanced Quantum Mechanics,* Addison-Wesley, 1967.
3. H. Feshbach and F. Villars, Rev. Mod. Phys. 30 (1958) 24.
4. Guang-jiong Ni, The relation between space-time inversion and particle-antiparticle transformation, Journal of Fudan University (Natural Science), 1974, No.3-4, 125-134, G-J Ni and S-q Chen, Internet, hep-th/9508069 (1995), to appear in a book *"Photon and Poincare Group"*, Edited by V. Dvoeglazov (NOVA Science Publishers, Inc. 1999).
5. G-j Ni and S-q Chen, On the essence of special relativity, ibid, 35 (3), (1996) 325-334.
6. Guang-jiong Ni and Su-qing Chen, Relativistic stationary Schrödinger equation for many-particle system, ibid, 36 (3) (1997)247-252, Guang-jiong Ni, hep.th/9708156.
7. J. J. Griffin, "An Alternative window upon the GSI sharp lepton problem" in *International Conference on Physics Since Parity Symmetry Breaking, in Memory of Professor C. S. Wu*, Editor F. Wang (World Scientific 1998), 537.
8. J. Schwinger, Proc. Nat. Acad. Sc USA, 44 (1958) 223.
9. E. J. Konopinski and H. M. Mahmand, Phys. Rev. 92 (1953) 1045.
10. E. C. G. Stüeckelberg, Helv. Phys. Acta 14 (1941) 32L, 588.
11. R. P. Feynman, Phys. Rev. 76 (1949) 749, 769.
12. G-j Ni and H. Guan, Einstein-Podolsky-Rosen Paradox and Antiparticles, Preprint, quant-ph/9901046.
13. G-j Ni, W-m Zhou, and J. Yan, The Klein paradox of Klein-Gordon equation and antiparticle, Preprint, quant-ph/9905044.
14. G-j Ni, To enjoy the morning flower in the evening——Is special relativity a classical theory? Kexue (Science) 50 (1) (1998) 29-33; Internet, quant-ph/9803034; To enjoy the morning flower in the evening——Where is the subtlety of quantum mechanics? ibid, 50 (2) (1998) 38-42; Internet, quant-ph/9804013; To enjoy the morning flower in the evening——What does the appearance of infinity in physics imply? ibid, 50 (3) (1998) 36-40; Internet, quant-ph/9806009. The English version of these three papers will be published in a book "*Photon: Old Problems in Light of New Ideas*", Edited by

V. Dvoeglazov (NOVA Science Publishers, Inc.).

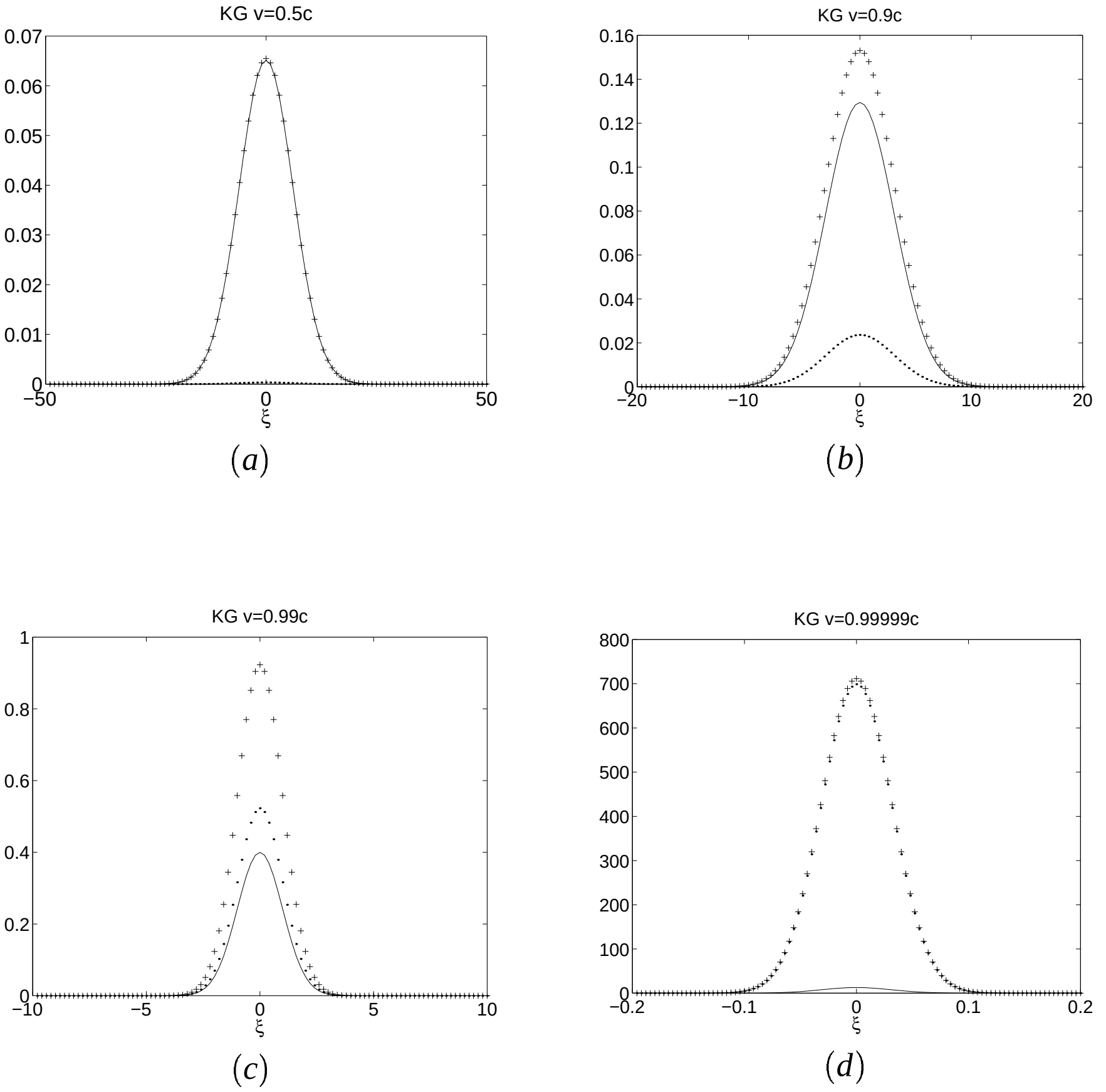

Figure 1: The wave packet of Klein-Gordon particle (e.g. $\pi^-$) for four velocities. (a) v=0.5c. (b) v=0.9c. (c) v=0.99c. (d) v=0.99999c. The +, ., and - curves denote the profiles of $|\varphi|^2$, $|\chi|^2$, and $\rho = |\varphi|^2 - |\chi|^2$ respectively. $\xi = m_0 c(z - vt)/\hbar$ is a dimensionless quantity.

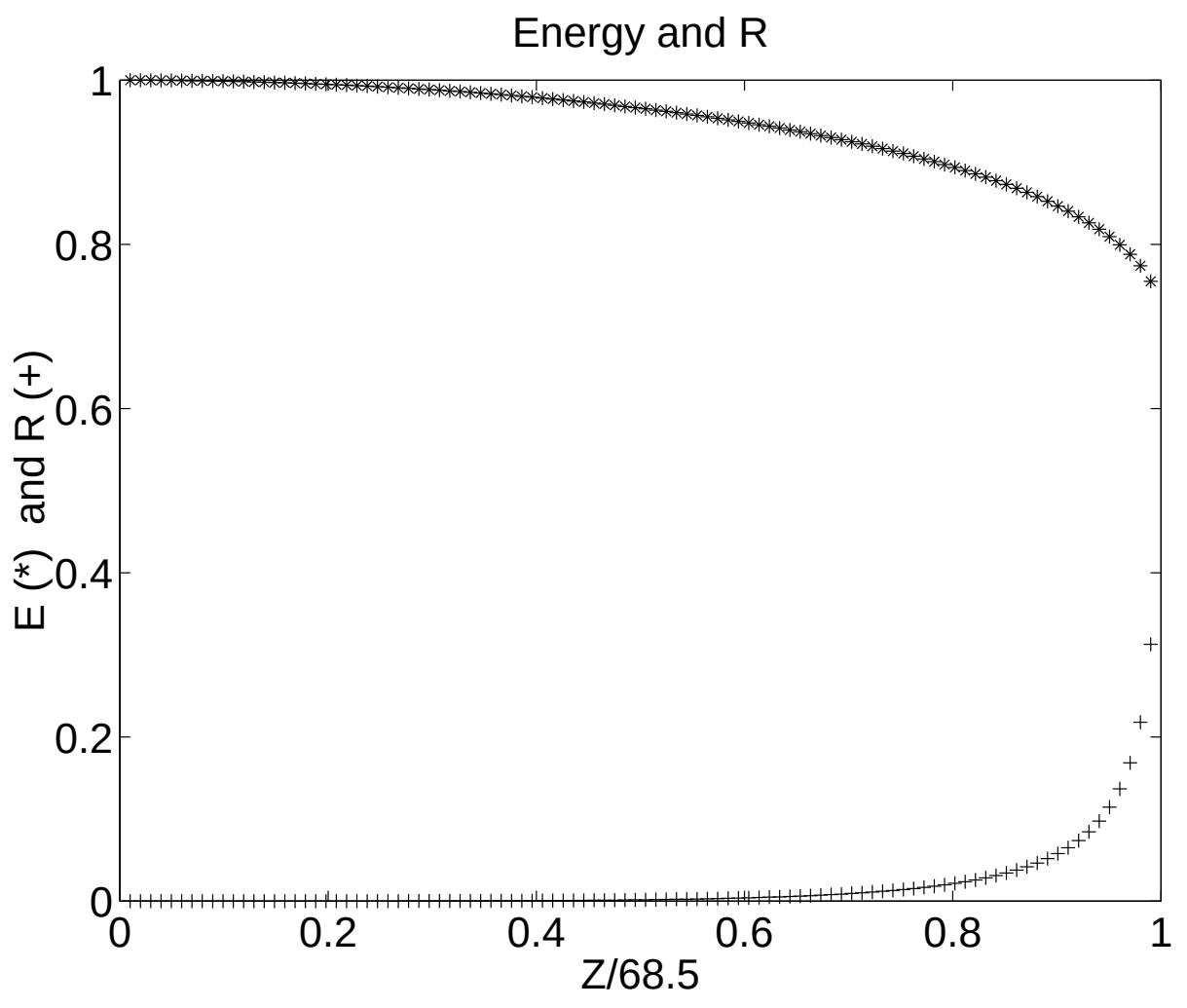

Figure 2: The * and + curve denote $E_{1S}^{KG}/m_0c^2$ and $R_{1S}^{KG}$ versus $Z/68.5$ respectively.

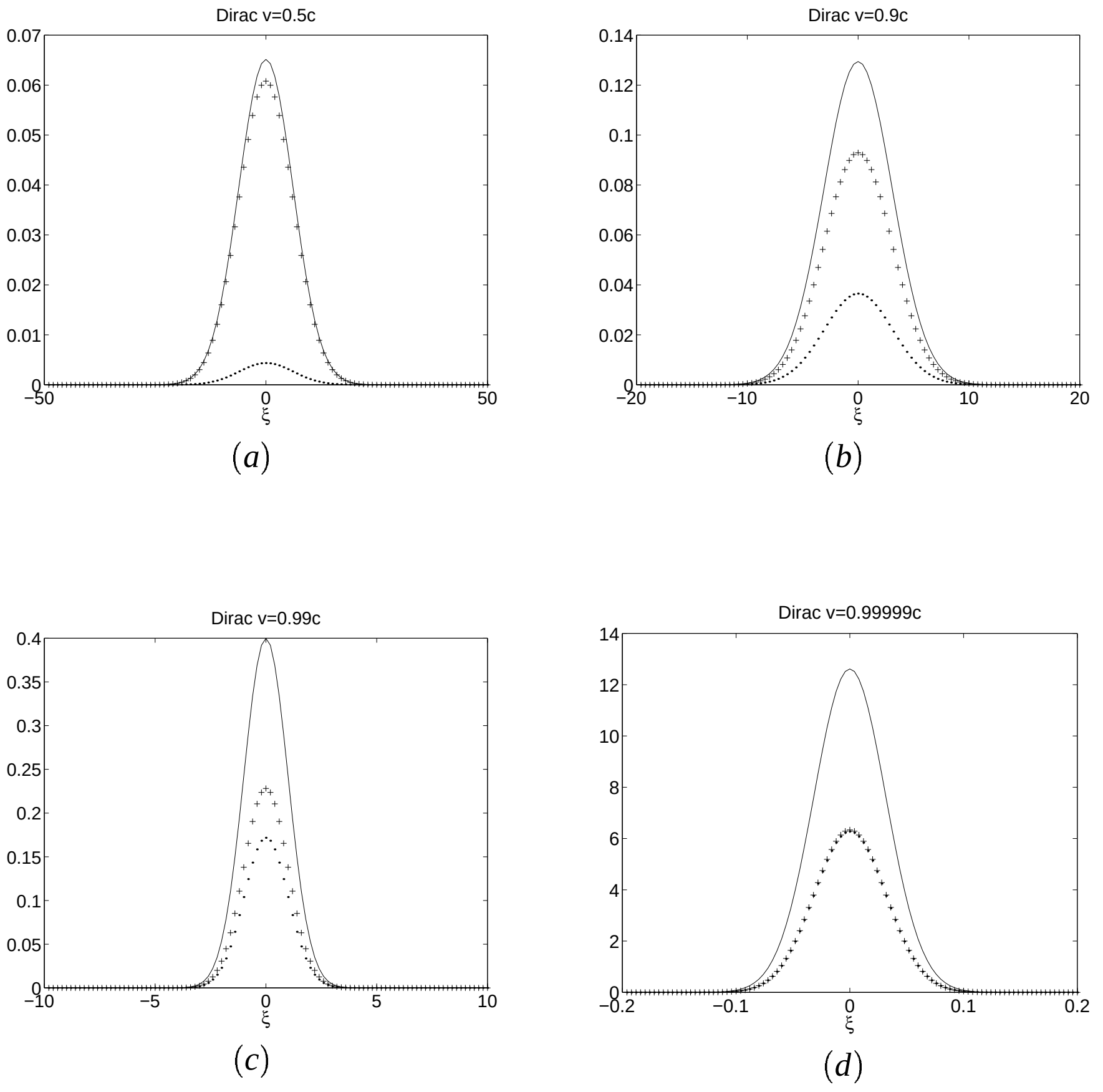

Figure 3: The wave packet of Dirac particle (e.g. the electron) for four velocities. (a) v=0.5c. (b) v=0.9c. (c) v=0.99c. (d) v=0.99999c. The +, ., and - curves denote the profiles of $\varphi^\dagger\varphi$, $\chi^\dagger\chi$, and $\rho = \varphi^\dagger\varphi + \chi^\dagger\chi$ respectively. $\xi = m_0 c(z - vt)/\hbar$ is a dimensionless quantity.

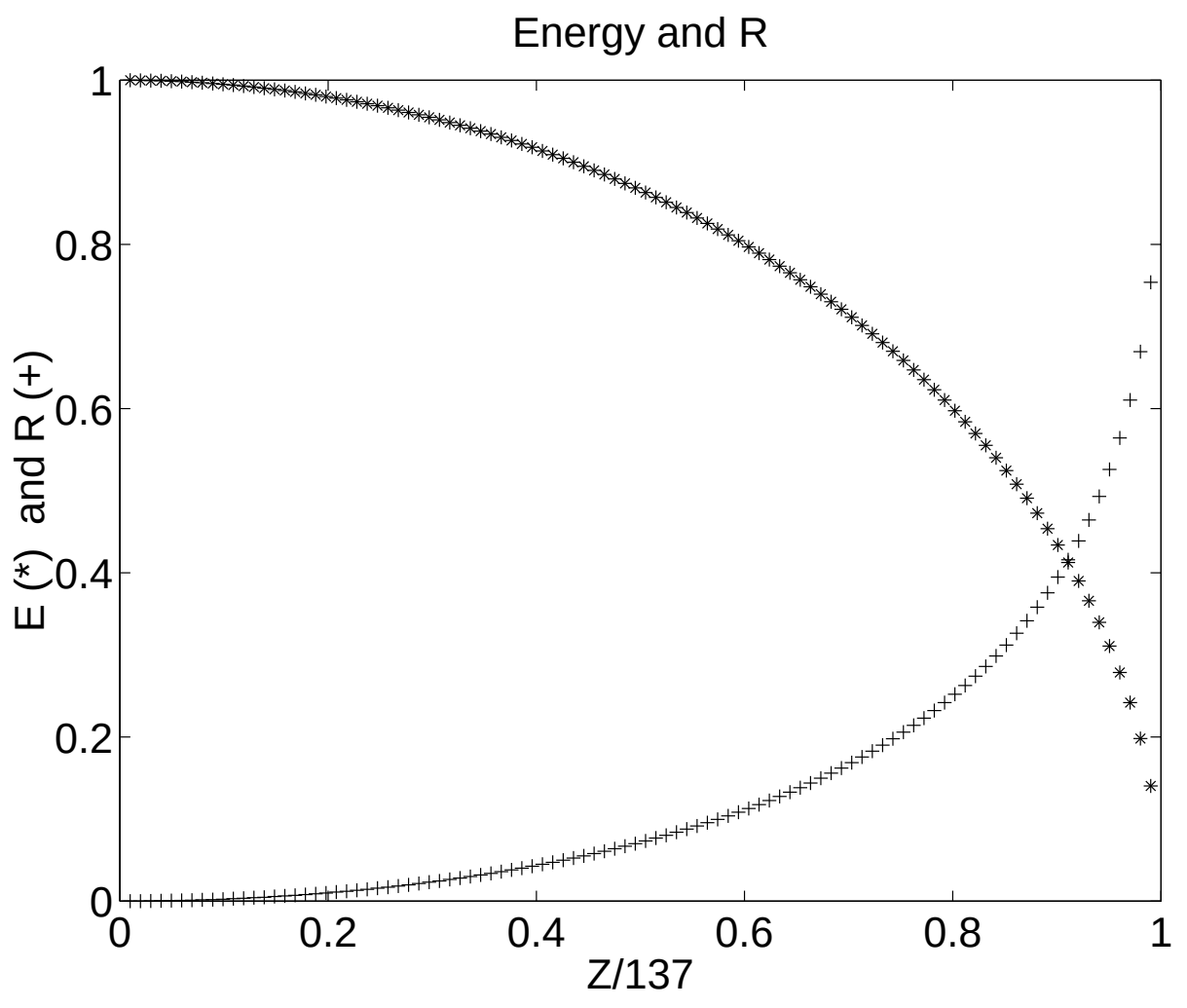

Figure 4: The * and + curve denote $E_{1S}^{Dirac}/m_0c^2$ and $R_{1S}^{Dirac}$ versus Z/137 respectively.